\documentclass[pra,aps,showpacs,superscriptaddress,twocolumn]{revtex4}
\usepackage{color} 
\usepackage{times}
\usepackage{amssymb} 
\usepackage{amsmath} 
\usepackage{ragged2e}
\usepackage{graphicx}
\usepackage{epstopdf}
\usepackage[english]{babel}
\usepackage{mathrsfs}
\usepackage{textcomp}
\usepackage{amsthm}
\usepackage{amsfonts}
\usepackage{hyperref}
\usepackage{soul}

\begin{document}

\title{Equivalence classes of Fibonacci lattices and their similarity properties}

\author{N. Lo Gullo}
\affiliation{Dipartimento di Fisica e Astronomia, Universit\`a degli Studi di Padova,
Padova, Italy}
\affiliation{CNISM, sezione di Padova}
\author{L. Vittadello}
\affiliation{Dipartimento di Fisica e Astronomia, Universit\`a degli Studi di Padova,
Padova, Italy}
\author{M. Bazzan}
\affiliation{Dipartimento di Fisica e Astronomia, Universit\`a degli Studi di Padova,
Padova, Italy}
\affiliation{CNISM, sezione di Padova}
\author{L. Dell'Anna}
\affiliation{Dipartimento di Fisica e Astronomia, Universit\`a degli Studi di Padova,
Padova, Italy}
\affiliation{CNISM, sezione di Padova}

\date{\today}
\begin{abstract}
We investigate, theoretically and experimentally, the properties of
Fibonacci lattices with arbitrary spacings. Differently from periodic
structures, the reciprocal lattice and the dynamical properties of
Fibonacci lattices depend strongly on the lenghts of their lattice
parameters, even if the sequence of long and short segment, the Fibonacci
string, is the same. In this work we show that, by exploiting a self-similarity
property of Fibonacci strings under a suitable composition rule, it
is possible to define equivalence classes of Fibonacci lattices. We
show that the diffraction patterns generated by Fibonacci lattices
belonging to the same equivalence class can be rescaled to a common
pattern of strong diffraction peaks thus giving to this classification
a precise meaning. Furthermore we show that, through the gap labeling
theorem, gaps in the energy spectra of Fibonacci crystals belonging
to the same class can be labeled by the same momenta (up to a proper
rescaling) and that the larger gaps correspond to the strong peaks
of the diffraction spectra. This observation makes the definition
of equivalence classes meaningful also for the spectral, and therefore
dynamical and thermodynamical properties of quasicrystals. Our results
apply to the more general class of quasiperiodic lattices for which
similarity under a suitable deflation rule is in order. 
\end{abstract}

\maketitle
\section{Introduction}
Since the first experimental proof of the existence of solids lacking
of translational invariance, but exhibiting a discrete Bragg diffraction
spectrum \cite{shechtman1984}, the study of \emph{quasicrystals}
attracted quite a lot of attention. The impact of this discovery on
the scientific community was such that in 1992 the former definition
of crystal had to be modified in order
to include those structures whose diffraction pattern witnesses long
range order yet lacking translational invariance \cite{levine1986,senechal1995,janot1994}.
More generally, the study of quasiperiodic geometries has been recently
the subject of different fields all devoted to the propagation of
waves through quasiperiodic potentials. The spectral properties of
quasicrystals have been recently used to engineer topological pumping
in optical waveguides \cite{kraus2012,kraus20122,verbin2013} and
in ultracold gases \cite{lohse2015,singh2015}. Engeneering of quasiperiodic
structures have also been employed in optical dielectric multilayers
for resonant transmission \cite{peng2004}, solar energy harvesting
\cite{lin2014}, plasmonics \cite{gopinath2008,huang2013} and nonlinear
optics \cite{lee2002,chou2014}. 

Dynamical and transport phenomena in this kind of structures are also
radically different compared to periodic media \cite{kimura2007,overy2015,li2013,dalnegro2014,radons2005}.
For usual periodic arrangements, dynamical and thermodynamical properties
are directly related, via the Bloch theorem, to the geometry of the
system. Quasiperiodic geometries, instead, lacks of translational
invariance so that a direct relation between their structure and their
dynamical properties is not generally known. It would be therefore
very interesting to find a sort of classification enabling one to
group together different aperiodic systems on the basis of some similarity
between their geometric arrangements. In this paper we attempt to
define such a classification, showing that quasiperiodic structures
whose geometry is related by a suitable mathematical transformation
share the main characteristics of their reciprocal lattice and of
their pseudo-band structure.

\section{Generalized Fibonacci lattices}
In one dimension (1D) the paradigm of a quasicrystal is the Fibonacci
lattice (FL). The FL is a 1D lattice whose adjacent points have distances
belonging to the set $\{L,S\}$, standing for Long and Short respectively,
which are arranged according to a given sequence. Such a lattice can
be constructed by means of the cut and project technique \cite{senechal1995,janot1994,buczek2005}
thus obtaining for the coordinates of points on the real line \cite{levine1986}
(in units of $S$):
\begin{equation}
x_{n}^{\eta}=n-1+\frac{1}{\eta}\left\lfloor \frac{n}{\tau}\right\rfloor \label{eq:fibscaled}
\end{equation}
where $n$ is a natural positive number,$\lfloor x\rfloor$ is the
integer part of $x$ and $\eta=S/(L-S)$.
The most common instance found in literature is obtained for $\eta=\tau=\left(\sqrt{5}+1\right)/2$,
the golden ratio. In this case the canonical FL (CFL) is obtained,
such that the lengths are (up to a simple rescaling): $L=1+1/\tau=\tau$,
and $S=1$. Nevertheless it is possible to construct Fibonacci lattices
with $\eta\neq\tau$ (see App. \ref{app:cutandproj}). 
The distances $\Delta_{n}=x_{n+1}^{\eta}-x_{n}^{\eta}$
are either $L=1+1/\eta$ or $S=1$ and they are arranged according
to the Fibonacci string (FS) $LSLLSLSLLSLLSLSLLSLSL\cdots$. The latter
is any word made of two letters, $L$ and $S$, obtained by means
of the substitution rule $S\rightarrow L$ and $L\rightarrow LS$
starting from the letter $L$. We notice that a FS itself is independent
on the parameter $\eta$ and it only depends on the factor $1/\tau$.

\begin{figure}
\includegraphics[width=8cm]{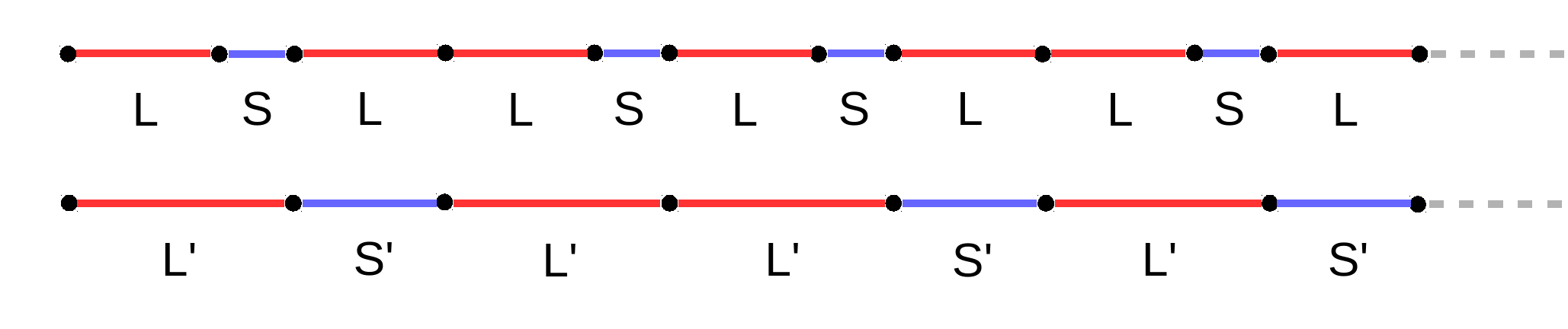}
\caption{The composition rules ($LS\rightarrow L'$, $L\rightarrow S'$) on
a semi-infinite Fibonacci lattice reproduces another Fibonacci lattice
with different lattice parameters.}

\label{fig:strings} 
\end{figure}

Conversely, given an infinite FS, a composition rule ($LS\rightarrow L'$
and $L\rightarrow S'$) can be defined such that the old and the new
strings are the same due to the peculiar properties of the Fibonacci
strings, as shown in Fig.~\ref{fig:strings}. For the special case
$\eta=\tau^{k}$, with $k$ a non-vanishing integer i.e. for the canonical
FL, this leads to a peculiar property: the new FL can be rescaled
to the original one. This case is the most commonly encountered in
literature, accompanied by the statement that the CFL is \emph{self-similar}.
It should be stressed however that this is not true for the general
case $\eta\neq\tau^{k}$. In this case, a non-canonical Fibonacci
lattice and the one obtained by applying the composition rule are
characterized by two different lenght ratios $\eta_{1}$ and $\eta_{2}$
because $L'/L\neq S'/S$. Therefore the new lattice cannot be transformed
into the old one by a simple rescaling.

We call $\mathcal{C}$ the operator corresponding to the effect of
the composition rule on the FL $x_{n}^{\eta}$. It is not difficult
to show that $\mathcal{C}\left(x_{n}^{\eta}\right)=\eta_{1}x_{n}^{\eta_{1}}$,
where $\eta_{1}=1+1/\eta$: the composition rule maps a FL $x_{n}^{\eta}$
into another FL, characterized by a new ratio $\eta_{1}$ and rescaled
by $\eta_{1}$ (see Fig. \ref{fig:strings}). If $\eta=\tau$ then
$\eta_{1}=\tau$ (recall that $\tau^{2}-\tau-1=0$) and therefore
the CFL is self-similar. If the composition rule is applied $k$ times,
the initial FL is mapped into 
\begin{equation}
\mathcal{C}^{(k)}(x_{n}^{\eta})=\left(\prod_{i=1}^{k}\eta_{i}\right)x_{n}^{\eta_{k}}=\left(F_{k+1}+\frac{F_{k}}{\eta}\right)x_{n}^{\eta_{k}}\label{eq:fibrinorm}
\end{equation}
where $\eta_{i}=1+1/\eta_{i-1}$ ($\eta_{0}=\eta$) and $F_{k}$ are
the Fibonacci numbers. With the help of this concept we define equivalence
classes for FLs by means of the following equivalence relation: \begin{justify}
\textbf{Definition -} \textit{Two Fibonacci lattices $x_{n}^{\eta_{a}}$
and $x_{n}^{\eta_{b}}$ are equivalent $(x_{n}^{\eta_{a}}\sim x_{n}^{\eta_{b}})$
if they are linked, up to a proper rescaling, by means of the composition
rule $\mathcal{C}$. The lattice with the minimum $\eta_{0}$ such
that $1/(\eta_{0}-1)$ is finite and positive is called the generator
of the equivalence class which is denoted by $[\eta_{0}]$.} \end{justify}
A simple way of labeling the elements of a given equivalence class
is by means of the continued fraction representation for the $\{\eta_{i}\}$.

Because of this, for quasiperiodic structures it is of limited practical
utility to talk about the support of the diffraction pattern. It is
more meaningful to describe the diffraction spectrum (and the reciprocal
lattice) in terms of the peaks which are significantly close to one,
which will be referred to as brightest peaks.
By means of the cut and project method outlined in App. \ref{app:cutandproj} 
it is possible to show (see App. \ref{app:diffpatt}) that the intensity $I(q,\eta)$ at points $q=Q(h,h')$ is
given by $\text{sinc}^{2}(Q_{\perp}(h,h')\Delta)$, where $Q_{\perp}(h,h')=2\pi d^{-1}(h(1+1/\eta)^{-1}-h')$
and $\Delta=\tau(\eta/(\eta+1))/2$. Therefore the brightest peaks
are found for pairs $(h,h')$ such that $Q_{\perp}(h,h')\approx0$
and thus for 
\begin{equation}
\frac{h}{h'}=1+\frac{1}{\eta}.\label{eq:condBPKs-1}
\end{equation}
Since $h$ and $h'$ are integers, the above condition can be satisfied
exactly only if $\eta$ is a rational number. On the other hand for
irrational $\eta$ we can resort to its continued fraction representation
in order to set the wanted precision to the above condition.

Let us now consider two FLs belonging to the same equivalence class
$x_{n}^{\eta_{0}}$ and $x_{n}^{\eta_{1}}$, with $\eta_{1}=1+1/\eta_{0}$.
By defining $h_{n}$ $(k_{n})$ and $h_{n}'$ $(k_{n}')$ as the numerator
and denominator of the $n$-th rational approximants of $1+1/\eta_{0}$
$(1+1/\eta_{1})$, the following relations hold: $k_{n}=h_{n}+h_{n}'$
and $k_{n}'=h_{n}$. The position of the brightest peaks of the FL
$x_{n}^{\eta_{1}}$ are then related to those of the FL $x_{n}^{\eta_{0}}$
by:
\begin{equation}
Q_{1}(k_{n},k_{n}')=\eta_{1}Q_{0}(h_{n},h_{n}')\label{eq:rescaled}
\end{equation}
 In other word, althought the two Fibonacci lattices $x_{n}^{\eta_{0}}$
and $\mathcal{C}(x_{n}^{\eta_{0}})=\eta_{1}x_{n}^{\eta_{1}}$ cannot
be rescaled one over the other (for the general case $\eta\neq\tau$),
their brightest peak pattern can, as a consequence of the fact that
they are related by the composition rule. Also the intensities of
the brightest peaks can be related as $I(\eta_{1}q,\eta_{1})\approx I(q,\eta_{0})+\frac{1}{\tau}(1-I(q,\eta_{0}))$
(for $q$ such that $I(q,\eta_{0})>0.5$) showing that the peaks of
the scaled lattice are even brighter than those of the original lattice.
This drives to the important conclusion that FL belonging to the same
equivalence class have diffraction spectra characterized by the same
pattern of brightest peaks, and are, in this sense, similar.

\section{Similarity of diffraction patterns}

To quantify the degree of similarity between the two spectra, we use
the Kullback-Leibler divergence (KLD), a quantity useful to compare
two distributions (normalized to unity over a common support). Let
us consider the diffraction spectra $I(q,\eta_{\alpha})$ and $I(q,\eta_{\beta})$
of two arbitrary FL's characterized by $\eta_{\alpha}\neq\eta_{\beta}$.
We define the normalized spectrum: $P(\nu q,\eta)=I(\nu q,\eta)/\int_{0}^{\infty}dkI(\nu k,\eta)$
where we introduced a scaling parameter $\nu$.

The KLD is defined as: 
\begin{equation}
D(\eta_{\alpha},\eta_{\beta},\nu)=\int_{0}^{\infty}dkP(k,\eta_{\alpha})\log\left(\frac{P(k,\eta_{\alpha})}{P\left(\nu k,\eta_{\beta}\right)}\right).
\end{equation}
By definition one has that the more similar the two diffraction spectra,
the smaller the value of the KLD. We will use it to measure if, for
given $\eta_{\alpha}$ and $\eta_{\beta}$, there exist a scaling
parameter $\nu$ for which the two spectra look similar. 

\begin{figure}
\includegraphics[width=8cm]{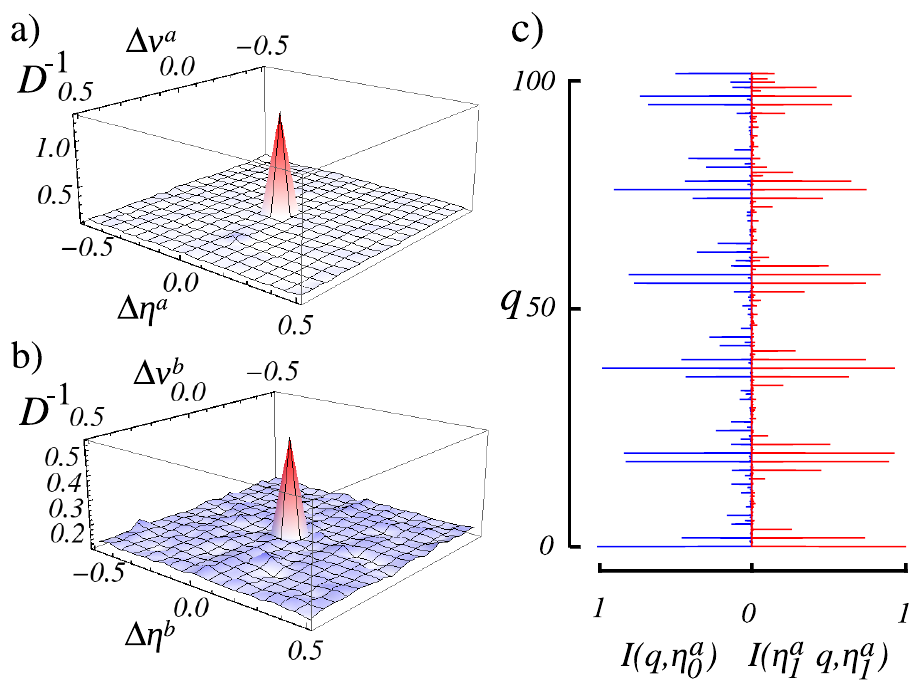}
\protect\caption{(Color online) Inverse of KL divergence $1/D(\eta_{0}^{a},\eta,\nu)$
comparing the diffraction spectra of the generator of a given class
$x_{n}^{\eta_{0}}$ with another Fibonacci lattice, for two choices
of $\eta_{0}$. {a)}$\eta_{0}^{a}=6/11$; {b)} $\eta_{0}^{b}=1/6$.
Here $\Delta\eta^{a}=\eta-\eta_{1}^{a}$, $\Delta\nu^{a}=\nu-\eta_{1}^{a}$
and $\Delta\eta^{b}=\eta-\eta_{1}^{b}$, $\Delta\nu^{b}=\nu-\eta_{1}^{b}\eta_{2}^{b}\eta_{3}^{b}$.
The maxima (minima of $D(\eta_{\alpha},\eta_{\beta},\nu)$) are obtained
at $\Delta\eta^{a,b}=0$ and $\Delta\nu^{a,b}=0$, indicating that
the two spectra with the higest degree of similarity corresponds to
lattices $\eta_{1}^{a}x_{n}^{\eta_{1}^{a}}=\mathcal{C}\left(x_{n}^{\eta_{0}^{a}}\right)$
and $\eta_{1}^{b}\eta_{2}^{b}\eta_{3}^{b}x_{n}^{\eta_{3}^{b}}=\mathcal{C}^{(3)}(x_{n}^{\eta_{0}^{b}})$
respectively, i.e. the first and the third element of the respective
equivalence classes. c) Direct comparison of the two diffraction spectra
for two FLs $x_{n}^{\eta_{0}^{a}}$ and $\eta_{1}^{a}x_{n}^{\eta_{1}^{a}}$.
The most prominent peaks of the diffraction pattern $I(q,\eta_{0}^{a})$
correspond with those of the (rescaled) spectrum $I(\eta_{1}^{a}q,\eta_{1}^{a})$. }

\label{fig:KLconfronto} 
\end{figure}
In Fig. \ref{fig:KLconfronto} a) and b) we plot $1/D(\eta_{0}^{\alpha},\eta,\nu)$
comparing two generators corresponding to $\eta_{0}^{a}=6/11$ and
$\eta_{0}^{b}=1/6$ with FLs obtained from them by applying the composition
rule $\mathcal{C}^{(n_{\alpha})}$ respectively $n_{a}=1$ and $n_{b}=3$
times. The intensities $I(q,\eta)$ are evaluated by means of eq.\ref{eq:theodiff-1}
for lattices with $N=300$ points. The maxima (minima of $D$) in
the two figures correspond to $(\eta,\nu)=(\eta_{1}^{a},\eta_{1}^{a})$
and $(\eta,\nu)=(\eta_{1}^{b},\eta_{1}^{b}\eta_{2}^{b}\eta_{3}^{b})$
respectively, in agreement with eq. (\ref{eq:rescaled}). This shows
that two FL's produce a similar diffraction pattern if and only if
they can be related via Eq. (\ref{eq:fibrinorm}) and therefore only
if they belong to the same equivalence class.

\begin{figure}
\includegraphics[width=8cm]{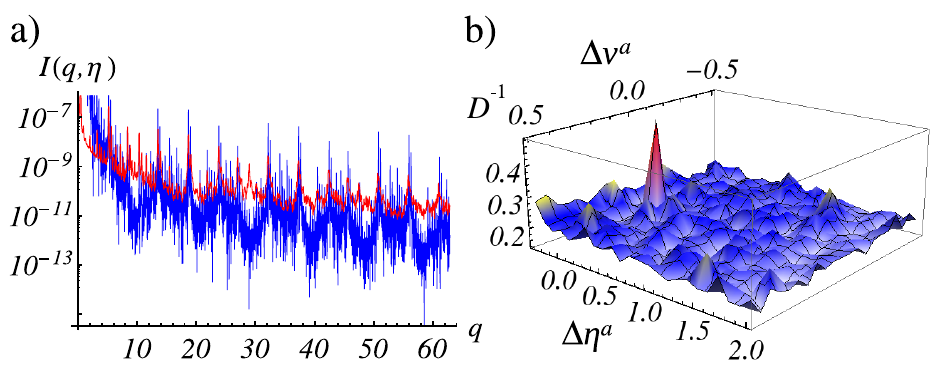}
\caption{(Color online). {a)} Comparison between the theoretical diffraction
pattern for the FL $x_{n}^{\eta_{0}^{a}}$ ($\eta_{0}^{a}=0.5454$,
solid blue curve) and the experimental one produced by a FL $\mathcal{C}\left(x_{n}^{\eta_{0}^{a}}\right)$
(solid red curve, $N=300$ lines and $L=23\;\mu m$ and $S=17\;\mu m$,
$\eta_{1}^{a}=2.8\overline{3}$). The theoretical spectrum has been
rescaled in accordance with Eq. (\ref{eq:rescaled}) and corrected
for the structure factor contribution. {b)} The inverse of KL divergence
$(D^{-1}(\eta_{1}^{a},\eta,\nu))$ between the (normalized over the
interval) experimental diffraction pattern and the theoretical diffraction
patterns for different generators and different scaling. Here $\Delta\eta^{a}=\eta-\eta_{0}^{a}$
and $\Delta\nu=\nu-1/\eta_{0}^{a}$. The maximum (minima of $D$)
is at $(\eta,\nu)=(\eta_{0}^{a},1/\eta_{0}^{a})$.}

\label{fig:scfib23_17} 
\end{figure}

In order to test our results on a real case, we performed a diffraction
experiment on two quasiperiodic diffraction gratings prepared using
a photorefractive direct laser writing (DLW) technique \cite{vittadello2015,argiolas2007}.
We used three gratings made up of $N=300$ lines all written in the
same substrate: (a) a periodic grating with spacing $L=23\mu m$;
two Fibonacci gratings with (b) $L=23\mu m$ and $S=17\mu m$ ($\eta_{1}^{a}=17/6$)
and (c) $L=23\mu m$ and $S=15\mu m$ ($\eta_{3}^{b}=15/8$) respectively.
So far we considered point lattices, but real structures are constitued
by some physical entity (basis) arranged on the points of our quasi-periodic
Fibonacci lattice. (For a detailed description of the experimental set up 
see App. \ref{app:expsetup}). 
For these cases, the diffraction pattern is given
by the sum in Eq.~(\ref{eq:theodiff-1}) multiplied by the square
modulus of a structure factor. The latter, in general,
does not posses any scaling property and therefore it is necessary
to correct for it when comparing different lattices. We did this experimentally
by using the data of the periodic grating to extract a phenomenological
expression for the structure factor as a function of $q$.
In Fig. \ref{fig:scfib23_17} {a)} we compare the experimental data
relative to the grating $\eta_{1}^{b}$ with the theoretical diffraction
pattern obtained from the generator of the corresponding equivalence
class, $\eta_{0}^{b}=1/6$. We observe that, once the spectra have
been rescaled in $q$ following eq. (\ref{eq:rescaled}) and corrected
in order to take into account the structure factor contribution to
the intensity of the peaks,
the most prominent diffraction features of the generator can be found
in the experimental data at the correct $q$ positions. The degree
of similarity between the spectrum of the generator and the experimental
one is confirmed by the KL divergence $D$ between the experimental
data points and the theoretical diffraction spectrum (with the inclusion
of the structure factor) calculated for a range of $\eta_{0}$ and
scaling factor $\nu$. In Fig. \ref{fig:scfib23_17} {b)} we show
it explicitly for the grating with $\eta_{1}^{a}$ and it is clear
that the maximum of $D^{-1}$ (minimum of $D$) is found at $\eta_{0}=\eta_{0}^{\alpha}$
and $\nu=\eta_{1}^{-1}$. Similar results are obtained for the grating
$\eta^{b}$.

\section{Energy spectra comparison}
\begin{figure}
\includegraphics[width=8cm]{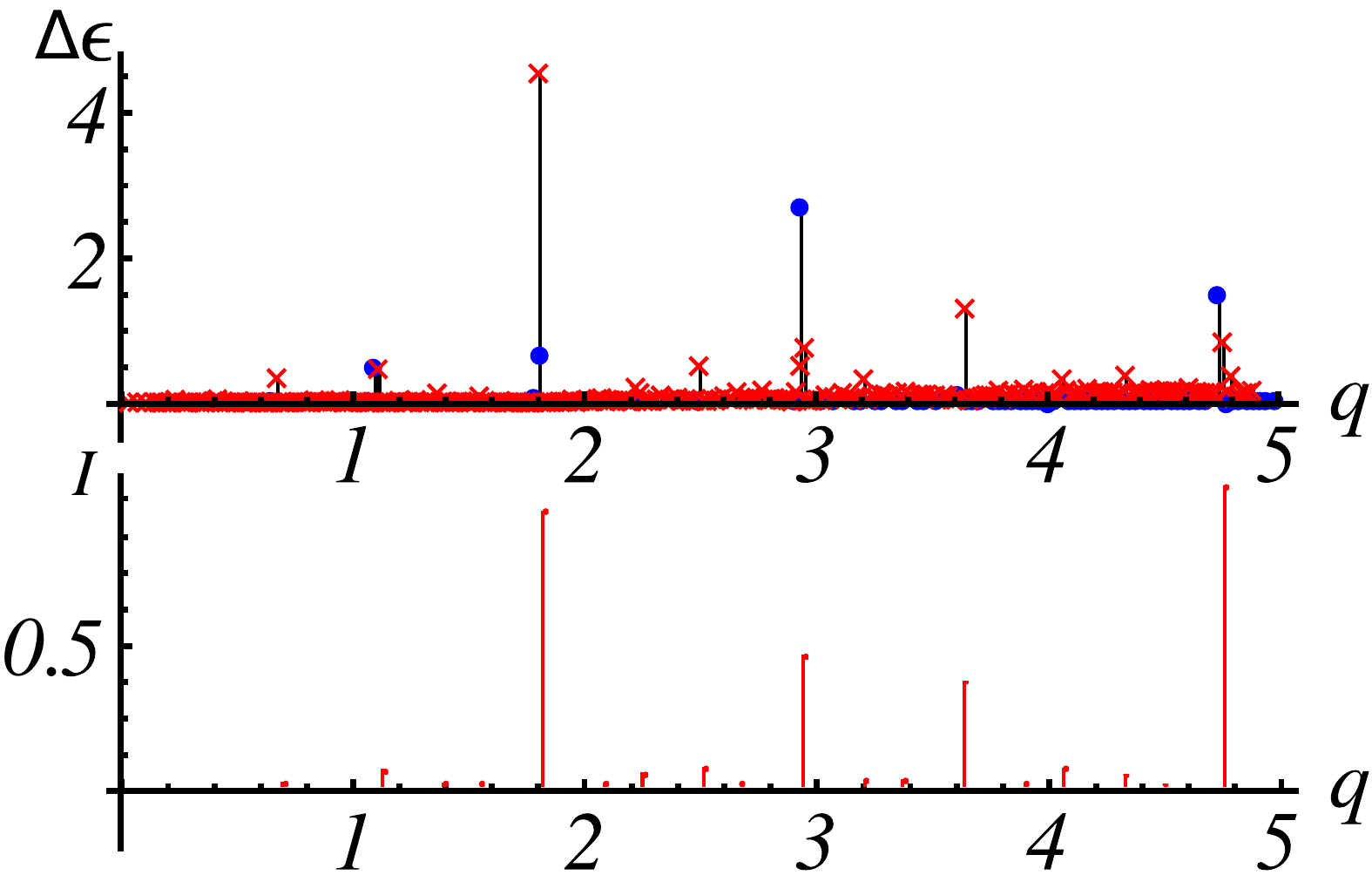}
\caption{(Color online). Plot of the energy level spacings of the Hamiltonian
in Eq.(\ref{eq:ham}) for a potential $V(x)$ having minima at the
points of FLs $x_{n}^{\eta}$ with (blue dots) $\eta=\eta_{0}^{a}$,
(red crosses) $\eta_{1}^{a}$. Below we compare the gaps with the
brightest peaks of $I(q/\eta_{1}^{a},\eta_{1}^{a})$.}

\label{fig:gaps} 
\end{figure}

We have therefore shown that all the FL belonging to the same equivalence
class have diffraction spectra that, although not equals, are characterized
by a similar pattern of bright peaks. This finding is of crucial importance
not only in scattering phenomena but also in transport ones. In fact
in a recent work \cite{vignolo2014} a method to unambiguously link
the gaps in the integrated density of states to the brightest peaks
in the diffraction pattern of the underlying potential has been proposed.
This is more general has it has been shown in a seminal paper by Luck \cite{luck1989}.
Let us consider for example the Hamiltonian for a particle in a 1D
lattice: 
\begin{eqnarray}
\hat{H} & = & -\frac{\hbar^{2}}{2}\frac{d^{2}}{dx^{2}}+V(x)\\
V(x) & = & -V_{0}\int dyf(x-y)\sum_{n}\delta(y-x_{n})\label{eq:ham}
\end{eqnarray}
where $x_{n}$ are the local minima of the potential and $f(x)$ is
introduced to account for the detailed shape of the potential minima
($V_{0}>0$). We will consider the case $x_{n}=x_{n}^{\eta}$ according
to the quasi-periodic sequence of Eq.~\ref{eq:fibscaled}. In Ref.~\cite{vignolo2014}
it has been shown that it is possible to label the energy gaps by
means of the brightest peaks of the diffraction spectrum. In particular
one has to consider the pseudo-momenta $q$ at which the square of
the Fourier transform of $V(x)$ acquires values greater than a given
threshold. This effectively corresponds to choose which free states
are effectively coupled by the potential and, therefore, where wider
gaps open in the single particle spectrum. One of the results presented
in Ref.~\cite{vignolo2014} is that this is equivalent to set a threshold
to the intensity of the peaks in the Bragg spectrum of the lattice.
This can be easily seen by considering the Fourier transform of the
potential $V(x)$, namely $V(q)=\int e^{\imath xq}V(x)dx$ whose square
modulus is given by: 
\begin{equation}
|V(q)|^{2}=S(q)I(q,\eta)
\end{equation}
where $S(q)$ is the square of the Fourier transform of $f(x)$ and
$I(q,\eta)$ is given by Eq. \ref{eq:theodiff-1}. It is clear from
what shown above and confermed by the experiment on the diffraction
patterns, that, apart from the contribution of the actual form of
the potential (which plays a role analogous to the structure factor
in diffraction experiments), the energy pseudo-band structure in reciprocal
space has the same shape (up to a rescaling) for the lattices beloging
to a given class. As an example, we computed the spectra of Eq.~(\ref{eq:ham})
in the case of Gaussian wells, namely $f(x)=e^{-x^{2}/2\sigma^{2}}$
for a system with $N=100$ minima and $x_{n}=x_{n}^{\eta}$ with $\eta=\eta_{0}^{a},\eta_{1}^{a}$
and $\eta=\eta_{0}^{b},\eta_{3}^{b}$. We choose $V_{0}=12$, $\sigma=0.1$.
In Fig.~\ref{fig:gaps} we plot the energy level spacing for lattices
characterized by $\eta_{0}^{a}=6/11$ (blue dots) and $\eta_{0}^{a}=17/6$
(red crosses) both belonging to the equivalence class $[\eta_{0}^{a}]$.
The momenta on the $x$-axis serve as a reference with respect to
the free particle dispersion relation $(\epsilon_{k}=k^{2}/2,V(x)=0)$
to show where the potential $V(x)$ opens the gaps. After rescaling
the momenta for the lattices with $\eta=\eta_{1}^{a}$ by $\nu_{a}=\eta_{1}^{a}$
we can clearly see that the gaps appear at the same points. On the
other hand these points correspond to the brightest peaks, where $I(q/\eta_{1}^{a},\eta_{1}^{a})$
calculated by Eq.~(\ref{eq:theodiff-1}) is sizeable. Similar results
are obtained (not shown) for the equivalence class $[\eta_{0}^{b}]$
with $\eta_{0}^{b}=1/6$ by considering the two lattices characterized
by $\eta_{0}^{b}$ and $\eta_{3}^{b}=15/8$, under the scaling $\nu_{b}=\eta_{1}^{b}\eta_{2}^{b}\eta_{3}^{b}$.

\section{Conclusions}
In conclusion, we investigated the diffraction spectra of FL's in
the general case $\eta=S/\left(L-S\right)\neq\tau$. We have shown
that it is possible to group different Fibonacci lattices into equivalence
classes whose elements share the main structural and dynamical properties
as witnessed by their diffraction spectra and the energy gaps. These
results show that the concept of equivalence classes for FLs has not
only a geometrical meaning but also an important role in the scattering,
dynamical and thermodynamical properties of the system, contained
in the energy spectrum. It is worth stressing once again that this
is a consequence of the self-similarity of FSs under the composition
rule and that FLs belonging to different equivalence classes cannot
be rescaled one over the other. The generator of a class is, in this
sense, the simplest structure giving a diffraction pattern which contains
the main features common to all of the other elements of the class.
Although we focused on the Fibonacci lattices, our arguments apply
to the more general class of quasicrystals for which deflation or
inflation rules can map the initial lattices into a similar ones.
\begin{acknowledgments}
NL and LD acknowledge financial support from MIUR, through FIRB Project
No. RBFR12NLNA\_002. LV and MB acknowledge financial support from
Università degli studi di Padova through \emph{Chip \& CIOP} project
No. CPDA120359. The authors are thankful to Prof. Camilla Ferrante
and Dr. Nicola Rossetto, from the Dipartimento di Chimica, Università
di Padova, for providing access to the direct laser writing setup. 
The authors thank J. Settino for providing the energy spectra 
of a particle in a Fibonacci like lattice.
\end{acknowledgments}

\appendix 
\section{Generalized Fibonacci lattices from cut and project method}
\label{app:cutandproj}
The Fibonacci lattices we considered in the main text 
can be constructed by means of the cut and project
technique. One possible construction has been presented in ref.\cite{buczek2005}
We prefer to resort to a more standard one and in what follows we will
generalize the one given in ref. \cite{senechal1995}.

\begin{figure}[h]
 \includegraphics[width=8cm]{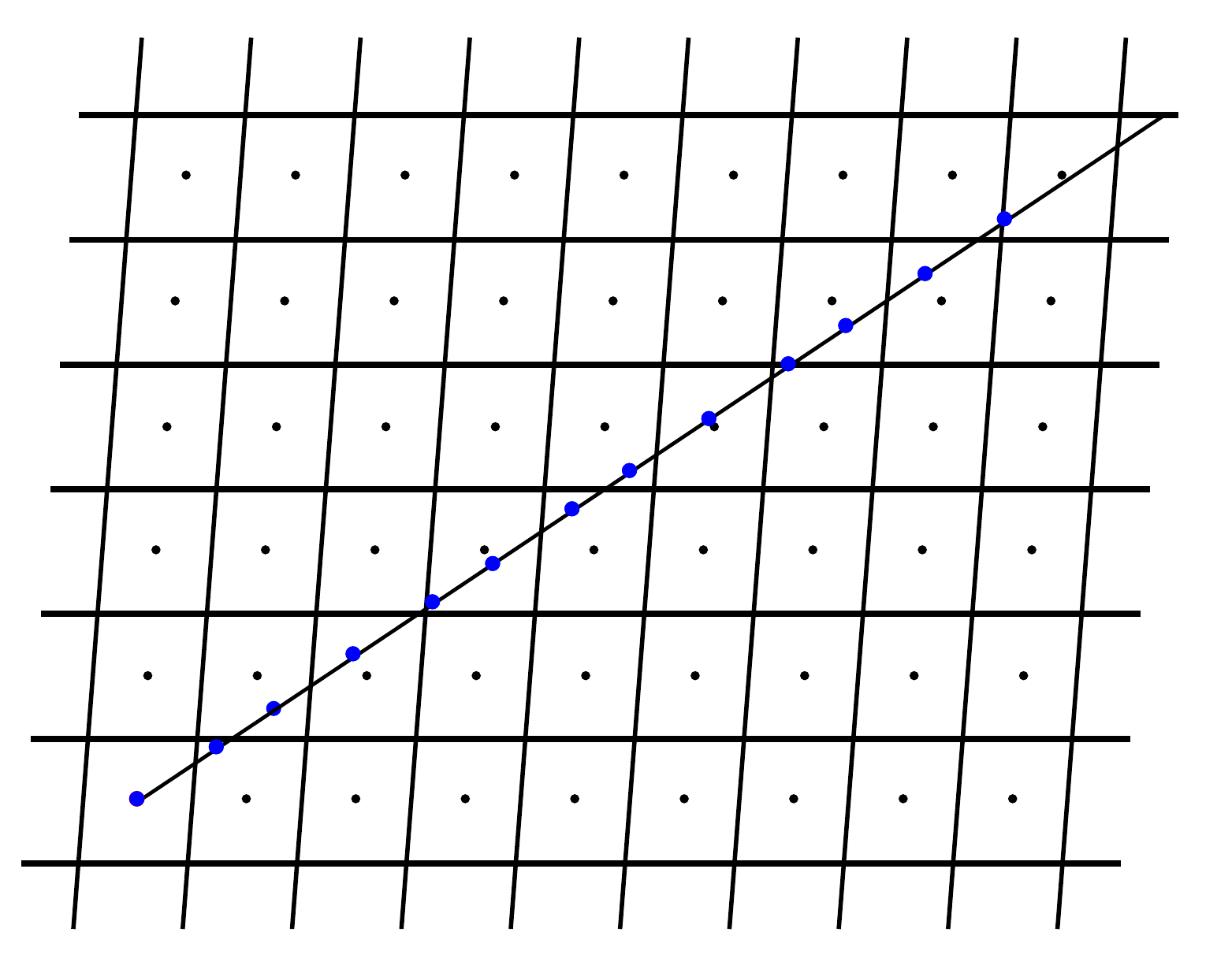}
 \caption{(Color online) Construction of a generalized Fibonacci lattice from a 2D periodic lattice
 by means of the cut and project method. Blue dots are the projection of points of the 2D lattice 
 whose Vonr\"oi cells are cut by the line $l_\tau$.}
 \label{fig:cap}
\end{figure}
Let us introduce a two-dimensional periodic lattice $\mathcal{I}_2^{p}$
and its lattice vectors ${\bf e}_1$ and ${\bf e}_2$ such that any point
of the lattice can be written as ${\bf p}=n_1{\bf e}_1+n_2{\bf e}_2$
with $n_1,n_2\in \mathbb{Z}$. 
Furthermore we introduce the line $l_\tau$ whose unit vector is 
$\hat l_\tau=(\cos(\theta_\tau),\sin(\theta_\tau))$ and 
the unit vector orthogonal to it $\hat l_\tau^{\perp}=(\sin(\theta_\tau),-\cos(\theta_\tau))$ 
such that $\tan(\theta_\tau)=\tau^{-1}$ where $\tau=(1+\sqrt{5})/2$.
The canonical Fibonacci lattice is constructed by projecting on the line $l_\tau$
the points of a square lattice (${\bf e}_1\cdot{\bf e}_2=0$, $|{\bf e}_1|=|{\bf e}_2|$) the points
whose Vonro\"i cell is cut by the line itself.
Let us notice that with this procedure the different points are unambiguously numbered
on the line $l_\tau$ once an origin and a direction have been chosen.
We are going to construct our Fibonacci lattices using this definition but allowing the 
two-dimensional lattice to be generic as in Fig.\ref{fig:cap}.
Nevertheless we will see that in order to obtain a Fibonacci lattice,
namely a one-dimensional set of points whose distances are distributed according to
the Fibonacci strings and with the wanted ratio between long and short segments,
we will have to restrict the set of the allowed two dimensional lattices.
Following the discussion in ref.\cite{senechal1995} in order for the line $\gamma \hat l_\tau$
to cut the Vonro\"i cell centered at point ${\bf p}$ it has to intersect 
the secondary diagonal of the cell (joining the northwest to southeast point of the cell).
The diagonals lie on lines parallel to $\delta ({\bf e}_1-{\bf e}_2)$
and whose points are given by $\delta ({\bf e}_1-{\bf e}_2)+n {\bf e}_1$ with $m\in\mathbb{Z}$.
Their intersection with the line $\gamma \hat l_\tau$ occurs at points 
$\left(\frac{n}{\tau},\frac{n}{\tau^2}\right)a$ where $a=(\tau^2/\sqrt{1+\tau^2}) A_{u.c.}/({\bf e}_1-{\bf e}_2)\cdot \hat l_{\tau}^{\perp}$ and $A_{u.c.}=({\bf e}_1\wedge{\bf e}_2)\cdot \hat z$ 
is the (oriented) area of the unit cell of the lattice.

\begin{figure}[h]
 \includegraphics[width=8cm]{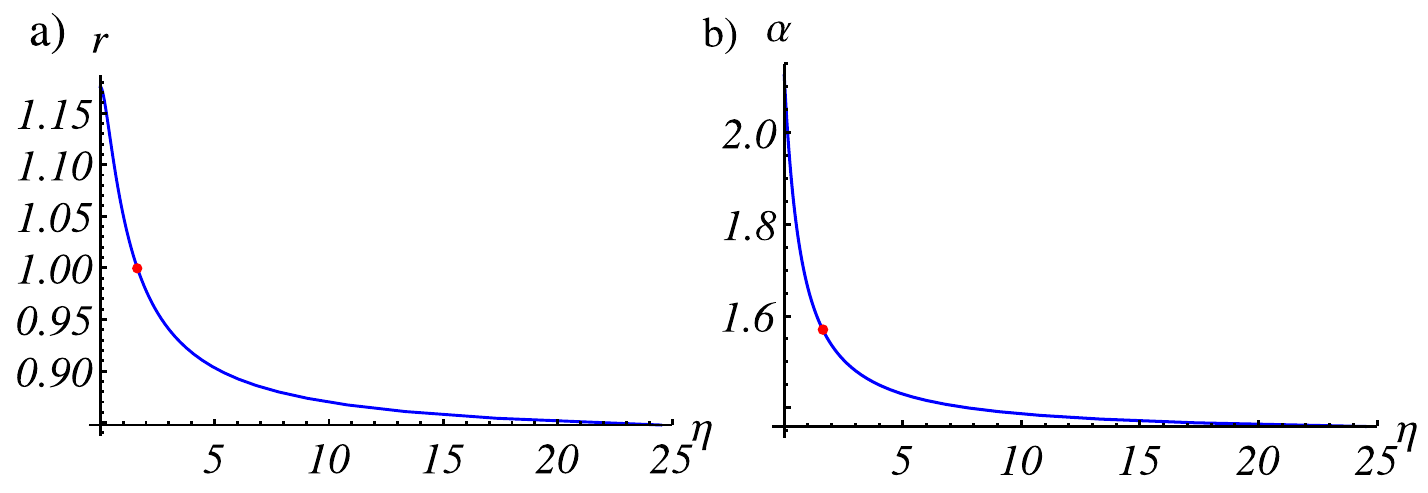}
 \caption{(Color online) {\bf a)} Values of $r=|{\bf e}_1|/|{\bf e}_2|$ as a function of $\eta$. 
 The red dot corresponds to the point $(\eta,r)=(\tau,1)$ for which the canonical Fibonacci lattice
 is obtained. {\bf b)} Values of the angle $\alpha$ between ${\bf e}_1$ and ${\bf e}_2$ as a function of $\eta$.
 The red dot corresponds to the point $(\eta,\alpha)=(\tau,\pi/2)$ for which the canonical Fibonacci lattice
 is obtained.
 }
 \label{fig:ralpha}
\end{figure}

This intersection points are inside the Vonro\"i cell centered at point $(u,v)$
if and only if
\begin{widetext}
\begin{eqnarray}
 u-\frac{|({\bf e}_1-{\bf e}_2)\cdot \hat x|}{2}<& n\; a\; \tau^{-1} &< u+\frac{|({\bf e}_1-{\bf e}_2)\cdot \hat x|}{2}\\
 v-\frac{|({\bf e}_1-{\bf e}_2)\cdot \hat y|}{2}<& n\; a\; \tau^{-2}&< v+\frac{|({\bf e}_1-{\bf e}_2)\cdot \hat y|}{2}
 \label{eq:fallvonr}
\end{eqnarray}
\end{widetext}
On the other hand each point of the lattice can be written as $n_1{\bf e}_1+n_2{\bf e}_2$
and $n_1+n_2=n$ because it is the $n$-th point to be projected.
Thus we can write $u=n_1 ({\bf e}_1-{\bf e}_2)\cdot \hat x + n {\bf e}_2\cdot \hat x$
and $v=n_1 ({\bf e}_1-{\bf e}_2)\cdot \hat y + n {\bf e}_2\cdot \hat y$
and above inequalities become
\begin{widetext}
\begin{eqnarray}
 \left(n_1-\frac{s_x}{2}\right)({\bf e}_1-{\bf e}_2)\cdot \hat x<& n\; (a\; \tau^{-1}-{\bf e}_2\cdot \hat x) &< \left(n_1+\frac{s_x}{2}\right)({\bf e}_1-{\bf e}_2)\cdot \hat x\\
  \left(n_1-\frac{s_y}{2}\right)({\bf e}_1-{\bf e}_2)\cdot \hat y<& n\; (a\; \tau^{-2}-{\bf e}_2\cdot \hat y) &< \left(n_1+\frac{s_y}{2}\right)({\bf e}_1-{\bf e}_2)\cdot \hat y
 \label{eq:fallvonr2}
\end{eqnarray}
\end{widetext}
where $s_x=\text{Sign}(({\bf e}_1-{\bf e}_2)\cdot \hat x)$ and similarly for $s_y$.
By means of the expression for $a$ it is easy to prove that $(a\; \tau^{-1}-{\bf e}_2\cdot \hat x)/({\bf e}_1-{\bf e}_2)\cdot \hat x=(a\; \tau^{-2}-{\bf e}_2\cdot \hat y)/({\bf e}_1-{\bf e}_2)\cdot \hat y$
and thus the two inequalities are equivalent to the inequality:

\begin{eqnarray}
 &&\left(n_1-\frac{1}{2}\right)< \frac{n}{\beta}< \left(n_1+\frac{1}{2}\right)\\
  &&\beta=1-\frac{{\bf e}_1\cdot\hat l_{\tau}^{\perp}}{{\bf e}_2\cdot\hat l_{\tau}^{\perp}}=1+r\frac{1}{\tau \sin(\alpha)-\cos(\alpha)}
 \label{eq:fallvonr3}
\end{eqnarray}
where $\cos(\alpha)={\bf e}_1\cdot {\bf e}_2/(|{\bf e}_1||{\bf e}_2|)$ and $r=|{\bf e}_1|/|{\bf e}_2|$.
Being $n_1$ an integer number the only possibility for the above inequalities to be satisfied is that $n_1=\lfloor \frac{n}{\beta}\rfloor$ where $\lfloor x\rfloor$ is the integer part of $x$.
After projecting onto $l_\tau$, the $n$-th point has coordinates on the the line $l_\tau$:

\begin{equation}
 x_n'=n \;{\bf e}_2\cdot\hat l_{\tau}+({\bf e}_1-{\bf e}_2)\cdot\hat l_{\tau}\;\left\lfloor \frac{n}{\beta}\right\rfloor
\end{equation}
By normalizing with respect to ${\bf e}_2\cdot\hat l_{\tau}$ 
we eventually obtain the one-dimensional lattice of points

\begin{eqnarray}
 x_n=n+\frac{1}{\eta}\;\left\lfloor \frac{n}{\beta}\right\rfloor\\
 \eta^{-1}=\left(\frac{{\bf e}_1\cdot\hat l_{\tau} }{{\bf e}_2\cdot\hat l_{\tau}}-1\right).
 \label{eq:1DFib}
\end{eqnarray}
In order for the above to be a Fibonacci lattice 
we require $\beta=\tau$ which is the case for
$\tau r=(\tau \sin(\alpha)-\cos(\alpha))$ and thus
$\eta=(\tau+\tan(\alpha))/((\tau-1)\tan(\alpha)-\tau^2)$.
Moreover we have to require that $r>0$ and $\eta>0$ which is the case 
for $\tan^{-1}(2\tau+1)<\alpha<\tan^{-1}(-\tau)+\pi$.
As it can be seen from figs.\ref{fig:ralpha} for any given $\eta>0$ 
there correspond a pair $(r,\alpha)$:

\begin{eqnarray}
\tan(\alpha)&=&\tau^2\frac{\eta \tau+1}{\eta-\tau}\\
r&=&\frac{(\tau \tan(\alpha)- 1)}{\tau \sqrt{1+\tan^2(\alpha)}}
\label{eq:randalpha}
\end{eqnarray}

and therefore a two dimensional lattice 
whose projection on the line $l_\tau$ returns the wanted FL:
\begin{eqnarray}
 x_n^{\eta}=n-1+\frac{1}{\eta}\;\left\lfloor \frac{n}{\tau}\right\rfloor\\
 \eta^{-1}=\left(\frac{{\bf e}_1\cdot\hat l_{\tau} }{{\bf e}_2\cdot\hat l_{\tau}}-1\right),
 \label{eq:1DFib}
\end{eqnarray}
where we shifted the whole lattice in order for the first point to have 
coordinate$x_1=0$ on the line $l_\tau$.
 
\section{Diffraction pattern}
\label{app:diffpatt}
We are interested in the calculation of the quantity:

\begin{equation}
 A(q_{\parallel})=\lim_{N\rightarrow\infty}\frac{1}{N} \sum\limits_{n}e^{\imath x_n^{\eta} q_{\parallel}}
\label{eq:I1D}
\end{equation}
where $x_n^{\eta}$  are given by eq.\ref{eq:1DFib}.
Using the unit vectors $\hat l_\tau$ and $\hat l_\tau^{\perp}$
we can write any point in space as $\vec r= x_{\parallel} \hat l_\tau+x_{\perp} \hat l_\tau^{\perp}$
and similarly for the variable $\vec q= q_{\parallel} \hat l_\tau+q_{\perp} \hat l_\tau^{\perp}$.
By introducing the quantity

\begin{equation}
 A_{X}(q_{\parallel},q_{\perp})=\lim_{N\rightarrow\infty}\frac{1}{N}\sum\limits_{n} e^{\imath \vec p_n\cdot \vec q}
\label{eq:Astrip2D}
\end{equation}
where we recall that $\vec p_n$ are points $\vec r_{n_1n_2}$ of the two dimensional periodic 
lattice which lie in a strip of width $2\Delta=|({\bf e}_1-{\bf e}_2)\cdot\hat l_{\tau}^{\perp}|$
around the line $\gamma \hat l_{\tau}$.
It is easy to see that $A(q_{\parallel})=A_{X}(q_{\parallel},0)$.
We therefore turn to the calculation of the latter.
By introducing the mass density of the two dimensional lattice $\rho(\vec r)=\sum_{m_1 m_2}\delta(\vec r-\vec r_{m_1m_2})$ and its Fourier transform $\rho(\vec r)=\int dk_{\perp}dk_{\parallel}e^{-\imath \vec r \cdot \vec k}\tilde\rho(\vec k)$ we can write:
\begin{eqnarray}
 A_{X}(q_{\parallel},q_{\perp})&=&\lim_{L\rightarrow\infty}\frac{1}{L} \frac{1}{2\Delta}\int dk_{\perp}dk_{\parallel}\\
 &&\int_{-\Delta}^{\Delta}dx_{\perp}\int_{-\infty}^{\infty} dx_{\parallel}\; e^{\imath \vec r\cdot (\vec q-\vec k)}\;\tilde\rho(\vec k)\nonumber
\label{eq:Astrip}
\end{eqnarray}
The above integrals can be calculated:
\begin{equation}
  A(q_{\parallel})=A_{X}(q_{\parallel},0)= \int dk_{\perp}\frac{\sin(k_{\perp}\Delta)}{k_{\perp}\Delta}\int dk_{\parallel}\;\tilde\rho(\vec k) \delta(k_{\parallel}-q_{\parallel})
\label{eq:Astrip2}
\end{equation}
which expresses the fact that the diffraction pattern of an infinite projected quasicrystal
is the convolution of the Dirac comb formed by the periodic higher dimensional periodic lattice
with the sinc function in the orthogonal space.

In particular $\tilde\rho(\vec k)$ is a Dirac comb peaked at points ${\bf k}_{hh'}=h{\bf w}_1+h'{\bf w}_2$
where we introduced the reciprocal lattice vectors for the dual of the 
two-dimensional periodic lattice $\mathcal{I}_2^{p}$:
\begin{eqnarray}
 {\bf w}_1=\frac{2\pi}{l_1}\left({\bf e}_1-{\bf e}_1\cdot \hat e_2 \hat e_2\right)\\
 {\bf w}_2=\frac{2\pi}{l_2}\left({\bf e}_2-{\bf e}_2\cdot \hat e_1 \hat e_1\right)\\
\end{eqnarray}
where $\hat e_{i}={\bf e}_{i}/|{\bf e}_{i}|$ and $l_1=|{\bf e}_{1}|^2-|{\bf e}_{1}\cdot\hat e_2|^2$
and similarly for $l_2$.
It is easy to check that ${\bf w}_i\cdot {\bf e}_j=2\pi \delta_{ij}$.
In what follows we assume that units are scaled such that ${\bf e}_2\cdot\hat l_{\tau}=1$.
In order to evaluate the parallel and perpendicular components of vectors 
belonging to the reciprocal space we need to evaluate ${\bf w}_i\cdot \hat l_{\tau}$
and ${\bf w}_i\cdot \hat l_{\tau}^{\perp}$.
In order to do so it is useful to rewrite the vectors ${\bf e}_i$
as linear combinations of $\hat l_{\tau}$ and $\hat l_{\tau}^{\perp}$
by means of the expressions for $\eta$, $\beta$ and the relation between $\tan\alpha$
and $\eta$.
We thus obtain:

\begin{eqnarray}
 {\bf e}_1&=&\left(1+\frac{1}{\eta}\right)\hat l_{\tau}+\frac{1}{\tau}\left(1+\frac{1}{\eta}\right)\hat l_{\tau}^{\perp}\\
 {\bf e}_2&=&\hat l_{\tau}-\left(1+\frac{1}{\eta}\right)\hat l_{\tau}^{\perp}
\end{eqnarray}

It is now easy to check that:
\begin{eqnarray}
 {\bf w}_1\cdot\hat l_\tau=\frac{2\pi}{d}&\;&{\bf w}_2\cdot\hat l_\tau=\frac{2\pi}{\tau d}\\
 {\bf w}_1\cdot\hat l_\tau^{\perp}=\frac{2\pi}{d\left(1+\frac{1}{\eta}\right)}&\;&{\bf w}_2\cdot\hat l_\tau^{\perp}=-\frac{2\pi}{d}
\end{eqnarray}
where $d=(\tau+1/\eta)$.
Therefore we can define:

\begin{eqnarray}
 Q(h,h')={\bf k}_{hh'}\cdot\hat l_\tau&=&\frac{2\pi}{d}\left(h+\frac{h'}{\tau}\right)\\
 Q_{\perp}(h,h')={\bf k}_{hh'}\cdot\hat l_\tau^{\perp}&=&\frac{2\pi}{d}\left(\frac{\eta\;h}{\eta+1}-h'\right)
\end{eqnarray}

By means of eq.\ref{eq:Astrip2} we can thus write the intensities
of the diffracted points as:

\begin{equation}
  I(q_{\parallel},\eta)=|A(q_{\parallel})|^2=\sum\limits_{h,h'}\frac{\sin^2(Q_{\perp}(h,h')\Delta)}{(Q_{\perp}(h,h')\Delta)^2}\delta(q_{\parallel}-Q(h,h'))
\label{eq:Istrip2sum}
\end{equation}
where $\Delta=\tau(1+1/\eta)/2$ and we introduce the explicit dependence
of the intensity on the parameter $\eta$ which characterizes the FL.
As it can be seen, the diffraction spectrum consists of a set of sharp peaks centered on a dense set of reciprocal lattice points, as by choosing the appropriate values of h and h', any q can be approximated with arbitrary precision. However, not all these peaks have the same intensity.

\begin{figure}[h]
 \includegraphics[width=8cm]{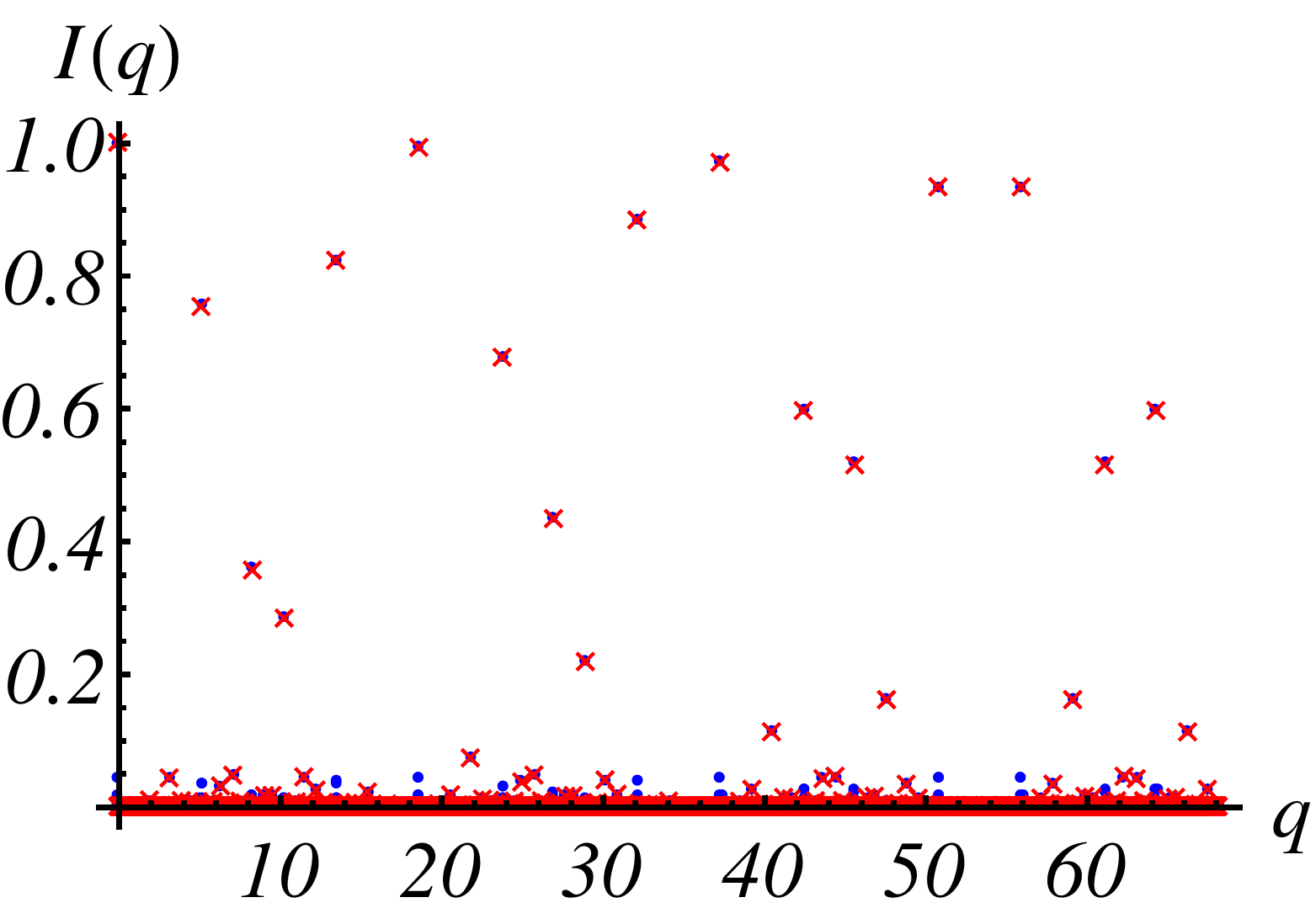}
 \caption{(Color online) Comparison between the values of $I(q_{\parallel})$ using the cut and project method and
 the direct evaluation in eq.\ref{eq:I1D} for a lattice of $N=300$ points and $\eta=17/6$. Blue dots are points corresponding to the value in eq.\ref{eq:Istrip2sum}
 whereas red cross are given by eq.\ref{eq:I1D}.}
 \label{fig:compsincsum}
\end{figure}

In fig.\ref{fig:compsincsum} we plot $I(q_{\parallel})$ as given by expression in
eq.\ref{eq:Istrip2sum} and its expression calculated explicitly by its definition
eq.\ref{eq:I1D} for a lattice of $N=300$ points and $\eta=17/6$.
We can see that as expected the peaks' intensities are well captured by eq.\ref{eq:Istrip2sum}
even for finite systems especially for peaks characterized by a significant intensity (>0.2).

We now consider the (Fraunhofer) diffraction pattern of a FL $x_{n}^{\eta}$:
\begin{equation}
I(q,\eta)=\lim_{N\rightarrow\infty}\frac{1}{N^{2}}\left|\sum\limits _{n}e^{\imath x_{n}^{\eta}\;q}\right|^{2}.\label{eq:theodiff-1}
\end{equation}
This quantity is important because it gives a direct experimental
access to the reciprocal lattice of our structure. We shall see that
this quantity is also encountered in the determination of a (pseudo)
energy dispersion relation~\cite{vignolo2014}. In the case of FLs
the values of $q$ at which a non-vanishing intensity is expected
are given by: 
\begin{equation}
Q(h,h')=\frac{2\pi}{d}\left(h+\frac{h'}{\tau}\right),\label{eq:reclatt}
\end{equation}
where $d=(\tau+1/\eta)$. By properly choosing the integers $h$ and
$h'$, any real number can be arbitrarily well approximated, showing that the reciprocal
lattice of a FL is dense in $\mathbb{R}$, contrarily to periodic
lattices which exhibit a discrete reciprocal lattice. Moreover it
can be shown \cite{senechal1995} that the diffraction pattern has
\textit{only} pure point support, lacking of a continuous part (according
to the classification of positive measures in the Lebesgue classification).

\section{Brightest peaks}
\subsection{Condition for brightest peaks}

In order to find the set of points in reciprocal
space corresponding to a strong diffracted intensity for a FL, the
following condition on the argument of the exponential in the expression
for the diffraction pattern has to hold:

\begin{equation}
 Q_{\perp}(h,h')\approx 0
\end{equation}
which is satisfied for

\begin{equation}
\frac{h}{h'}\approx 1+\frac{1}{\eta}.
\label{eq:condBPKs}
\end{equation}

Let us now consider the equivalence class $[\eta_{0}]$ and in particular
the sequence $x_{n}^{\eta_{0}}$. 
We can write $\eta_0$ in the continued fraction representation:
\begin{equation}
\eta_{0}=a_0+\frac{1}{a_1+\frac{1}{a_2+\frac{1}{a_3+...}}}\equiv[a_{0},a_{1},a_2,a_3,\cdots]
\end{equation}
For a rational number the sequence of numbers $a_i$ is finite, namely $\eta_0=[a_{0},a_{1},\cdots,a_n]$. 
On the other hand, if $\eta_0$ is irrational, it is possible to find a rational approximation
within the wanted error by increasing the number of terms in its continued fraction representation. 
It is easy to see that $\eta_{1}=1+1/\eta_{0}=[1,a_{0},a_{1},\cdots]$
and in general 
\begin{equation}
\eta_{k}=[\underbrace{1,1,\cdots,1}_{k},a_{0},a_{1},\cdots].
\end{equation}
With this notation is straightforward
to see that regardless of the value of the generator $\eta_{0}$,
the sequences of an equivalence class will tend to a Fibonacci sequence 
since $\lim_{k\rightarrow\infty}\eta_{k}=
[1,1,1,1,\cdots]=\tau$ 
\cite{limit}.
Using the continued fraction notation
we can write a sequence of rational approximants to $\eta_{0}$
as $a_{0}$, $\frac{a_{1}a_{0}+1}{a_{1}}$, $\frac{a_{2}(a_{1}a_{0}+1)+a_{0}}{a_{2}a_{1}+1}$, $\cdots$. 
Since both $h$ and $h'$ have to be integers the above 
condition~\ref{eq:condBPKs} is satisfied if we choose $h=s_n+t_n$ and $h'=s_n$
where $s_n$ and $t_n$ are the $n$-th approximants of $\eta_0$
namely $\eta_0\approx s_n/t_n$ and can easily be derived from the 
continued fraction representation of $\eta_0$.

It is worth stressing that if $\eta_{0}$ is a rational number ($\eta_{0}=a/b$, with $a,b\in\mathbb{N}$)
$h$ and $h'$ can be chosen such that $h/h'=(a+b)/a$. 
Thus, at points $q_m=Q(m(a+b),mb)=2m\pi$ ($m\in\mathbb{Z}$)
we have that $I(q_m,\eta_{0})=1$.
On the other hand, for irrational $\eta_{0}$ the condition is never satisfied exactly 
but we can resort to the rational approximants
of $\eta_{0}$ to estimate the positions at which the brightest peaks
appear.

\subsection{Relation between positions of brightest peaks}

Let $x_{n}^{\eta_{0}}$ and $x_{n}^{\eta_{1}}$ be two Fibonacci lattices 
belonging to the same equivalence class and their associated reciprocal lattices 
$Q_{0}(h,h')=2\pi d_{0}^{-1}(h+h'/\tau)$,
$Q_{1}(h,h')=2\pi d_{1}^{-1}(k+k'/\tau)$ respectively, where $d_{i}=\tau+1/\eta_{i}$.
By defining $h_{n}\;\;(k_{n})$ and $h_{n}'\;\;(k_{n}')$ as the numerator
and denominator of the $n-th$ rational approximants of $1+1/\eta_{0}\;\;(1+1/\eta_{1})$,
the following relations hold true $k_{n}=h_{n}+h_{n}'$ and $k_{n}'=h_{n}$.
By inserting these relations into the expression for $Q_{1}(k,k')$
we get:

\begin{eqnarray}
Q_{1}(k_{n},k_{n}') & = & \frac{2\pi}{d_{1}}\left(k_{n}+\frac{k_{n}'}{\tau}\right)\label{eq:newbrightspot}
   = \frac{2\pi}{d_{1}}\left(h_{n}+h_{n}'+\frac{h_{n}}{\tau}\right)\nonumber \\
 & = & \frac{2\pi\tau}{d_{1}}\left(h_{n}+\frac{h_{n}'}{\tau}\right)
   = \frac{d_{0}\tau}{d_{1}}\frac{2\pi}{d_{0}}\left(h_{n}+\frac{h_{n}'}{\tau}\right)\nonumber \\
 & = & \frac{d_{0}\tau}{d_{1}}Q_{0}(h_{n},h_{n}')=\eta_{1}Q_{0}(h_{n},h_{n}')\nonumber 
\end{eqnarray}
where in the last line we used the fact that $d_{0}\tau/d_{1}=\eta_{1}$.
This means that the lattice obtained by applying the composition rule
$\mathcal{C}(x_{n}^{\eta_{0}})=\eta_{1}x_{n}^{\eta_{1}}$
has brightest peaks at the same positions of the original lattice
only rescaled by a factor $eta_1$.

\subsection{Relation between intensities of brightest peaks}
From eq.\ref{eq:Istrip2sum} we can also estimate the relation between 
the intensities of the brightest peaks in the diffraction spectrum of two FL belonging to the same class.
Using the expression in eq.\ref{eq:Istrip2sum} and assuming $k_{\perp}\Delta\approx 0$
we can write $\sin^2(k_{\perp}\Delta)/(k_{\perp}\Delta)^2-1\approx (k_{\perp}\Delta)^2/9$.
Using the condition for $k_\perp\approx 0$ and following a calculation similar 
to that to determined relation between the positions of the brightest peaks
we find that $(k_{\perp}^1\Delta_1)^2=(k_{\perp}^0\Delta_0)^2/\tau^2$.
Therefore we have 
\begin{equation}
I(\eta_1 q,\eta_1)\approx I(q,\eta_0)+\frac{1}{\tau} (1-I(q,\eta_0)),
\end{equation}
meaning that the intensities of brightest peaks of the scaled lattice
are more intense of those of the original lattice by a term proportional to 
the difference between the maximum attainable intensity and the intensity of the 
original lattice intensities.

\begin{figure}[h]
 \includegraphics[width=8cm]{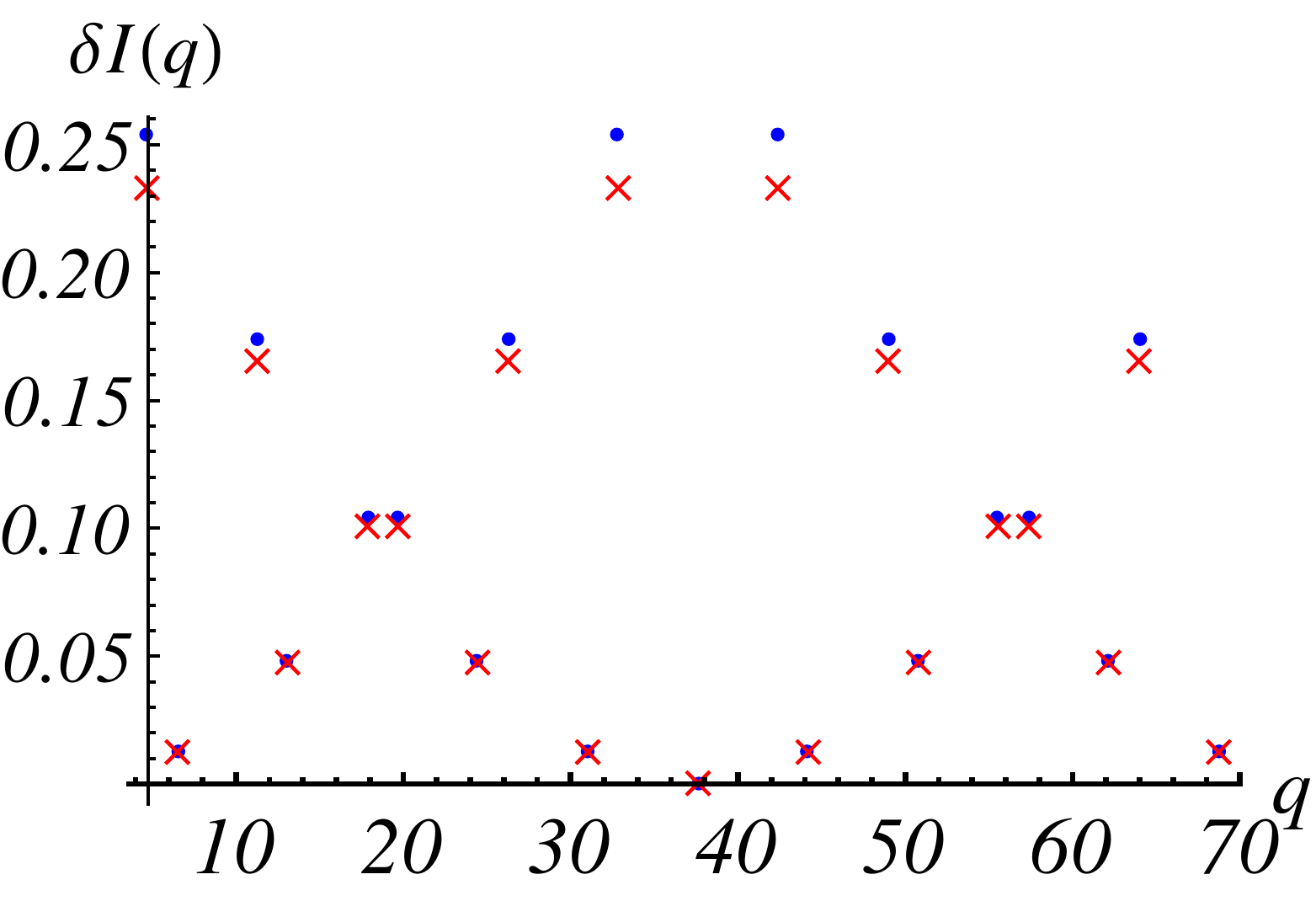}
 \caption{(Color online) Plot of (blue dots) $\tau^{-1} (1-I(q,\eta_0))$ and (red cross) 
 $(I(\eta_1 q,\eta_1)-I(q,\eta_0))$ for those $q$ for which $I(q,\eta_0)>0.5$ showing that the intensities of 
 the brightest peaks of the diffraction pattern of a FL $x_{n}^{\eta_1}$ 
 generated from a FL $x_{n}^{\eta_0}$ by means of the composition rule are more intense of those
 of the original lattice by a factor of $\tau$.}
 \label{fig:diffint}
\end{figure}

In fig.\ref{fig:diffint} we plot the quantities (blue dots) 
$\tau^{-1} (1-I(q,\eta_0))$ and (red cross) $(I(\eta_1 q,\eta_1)-I(q,\eta_0))$
for $q$ such that $I(q,\eta_0)>0.5$ and for lattices of $N=300$ sites and $\eta_0=6/11$
and $\eta_1=1+1/\eta_0$ respectively.

\section{Experimental setup}

To test experimentally the diffraction from FL's, a series of quasi-periodic 
diffraction gratings have been prepared using a photorefractive
direct laser writing (DLW) technique \cite{vittadello2015}. This technique
consists in scanning with a focused laser beam a photorefractive sample,
engraving on it a series of lines with a modified refractive index
with respect to the rest of the sample. The scanning movement is performed
by translating the sample with the aid of a computer-controlled
XY stage at constant speed of 50 $\mu m/s$. The nominal precision
of the translation stage is 0.5$\mu m$ for the conditions used in
this experiment. A frequency doubled diode pumped $\textrm{Nd:YWO}_{4}$
solid state laser (Coherent Verdi V5) emitting a CW beam at 532 nm
has been used as light source for DLW. The beam was suitably attenuated
by a series of neutral density filters and sent to a focusing microscope
objective (Olympus 100X/0.80) so that the power after the objective
was set at 17 mW. The substrate used to engrave the optical structures
is a slab of photorefractive lithium niobate doped with iron at the
nominal concentration of 0.1 mol\% in the melt. The sample was X-cut
with dimensions ($\textrm{X}\times\textrm{Y}\times\textrm{Z}$) $1\textrm{mm}\times8\textrm{mm}\times13\textrm{mm}$
and the lines were written on the X face, by scanning along the Y
direction with an ordinarily polarized beam. This process can induce
extraordinary refractive index changes as large as $10^{-3}$in the
written lines and therefore can be used to produce arbitrary diffraction
structures. 
The diffraction pattern of these structures was measured with the
help of a computer-controlled optical diffractometer in which the
sample and the detector were mounted on two co-axial goniometers
that were independently controlled by a computer \cite{argiolas2007}.
An optical beam produced by a He-Ne laser at 632.8 nm with a power
of 4 mW was expanded, polarized along the extraordinary direction
and finally transmitted through the sample surface, resulting in a
clearly visible diffraction pattern. This pattern was measured by
a Si photodiode and a lock-in amplifier and recorded on the computer
as a function of the detector and of the sample angle.

\section{Experimental diffraction patterns}
\label{app:expsetup}
\subsection{Structure factor}
In order to compare the experimental data with the theoretical calculation
we need to take into account that our gratings are made up of a (quasi)
periodic repetition of a region with a modified refractive index,
$\Delta n(x)$. This leads to the fact, well known from standard diffraction
theory, that the diffracted intensity in reciprocal space is proportional
to the product of two terms: a first one, $S(q)$=$\left|\int\Delta n(x)e^{\imath xq}dx\right|^{2}$
which depends on the detailed structure of the repeated unit of the
grating (\emph{structure factor}) and a second term due to lattice
geometry, which is the true object of this study:

\begin{equation}
I_{R}(q,\eta)=S(q)\frac{1}{N^2}\left|\sum\limits _{n}e^{\imath x_{n}^{\eta}q}\right|^{2}=S(q)I(q,\eta)\label{eq:realdiff}
\end{equation}

\begin{figure}[h]
\includegraphics[width=8cm]{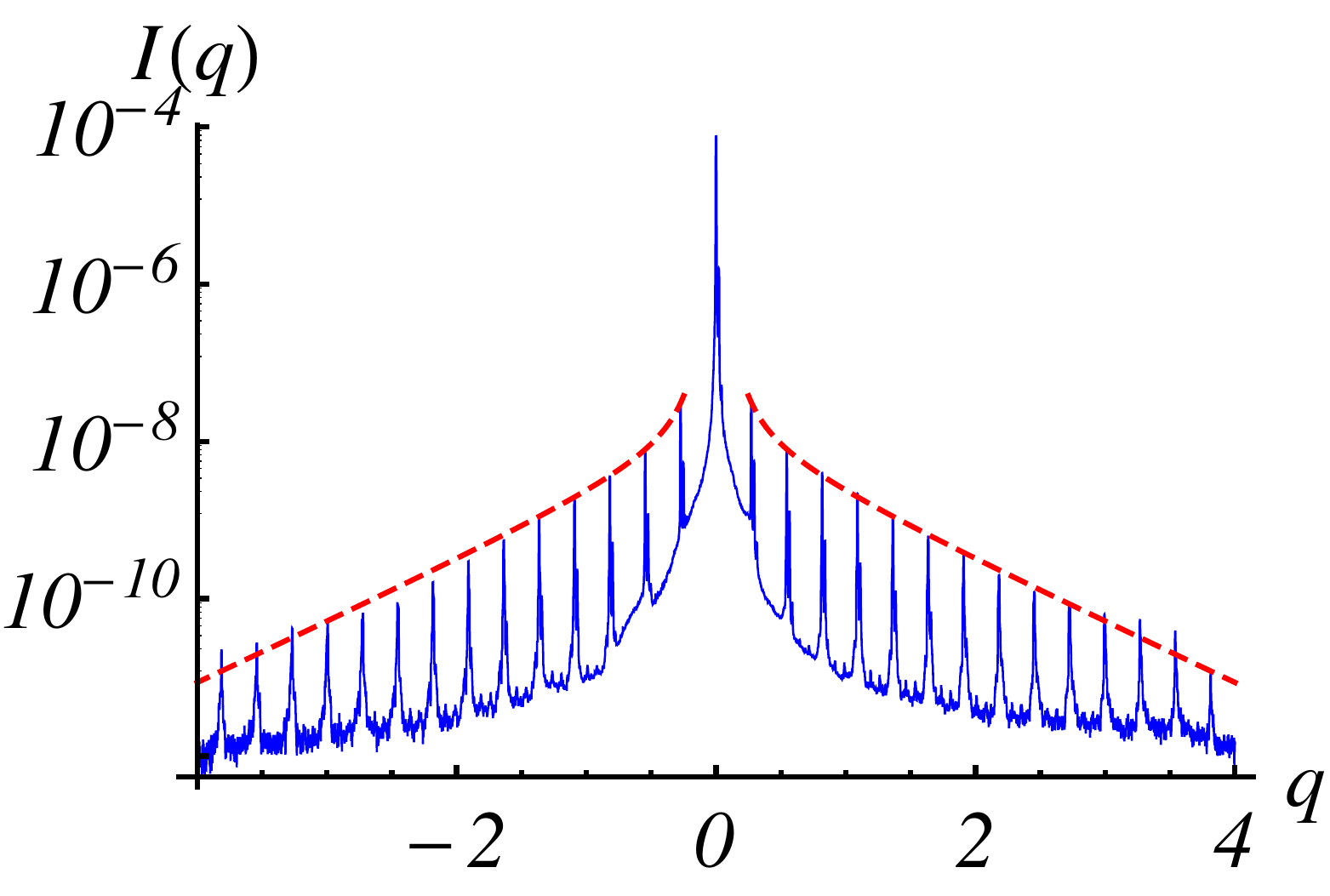}
\protect\protect\caption{(Color online). Experimental diffraction pattern of the periodic grating
with spacing $L=23\;\mu m$ (blue, solid curve) and fit of the satellite
peak intensities using function \ref{eq:strucfact} (red, dashed curve).
}
\label{fig:per23} 
\end{figure}

The structure factor modulates the intensity of the lattice diffraction
pattern, complicating the comparison between experiments and theory.
In principle $S(q)$ could be calculated by knowing the details of
the refractive index profile changes produced by our technique. Here
we used another approach which exploits the fact our samples differ
only for the line position sequence $x_{n}$. We can use therefore
the periodic grating (Fig. \ref{fig:per23}) to measure the function
$S(q)$ directly at the reciprocal lattice points $\{q_{M}^{i}\}$
of the periodic grating, where $I(q_{M}^{i})$ has local maxima. 

We found that the following phenomenological functional form for $S(q)$
describes adequately the peak intensity in the whole range of measured
values (see fig. \ref{fig:per23}):

\begin{equation}
S(q)=S_{0}e^{-\lambda q-\frac{q_{0}}{q}}\label{eq:strucfact}
\end{equation}
where the parameters $S_{0},\lambda,q_{0}$ are determined by a least
square fit in the range $q\in[0.5,3]\;\mu m^{-1}$ excluding the last
peaks because the corresponding momenta where comparable with a length scale
of the order of the optical waveguide width. We also notice
that all measured diffraction patterns drop almost to zero outside
the interval $q\in[-4,4]\;\mu m^{-1}$; this is due to the fact that
our lines have a width determined by the laser writing optics which
is not smaller than 2 $\mu m$, so that our diffraction pattern cannot
probe $|q|>3\;\mu m^{-1}$.

\subsection{Comparison with theoretical patterns}
\begin{figure}[h]
\includegraphics[width=7cm]{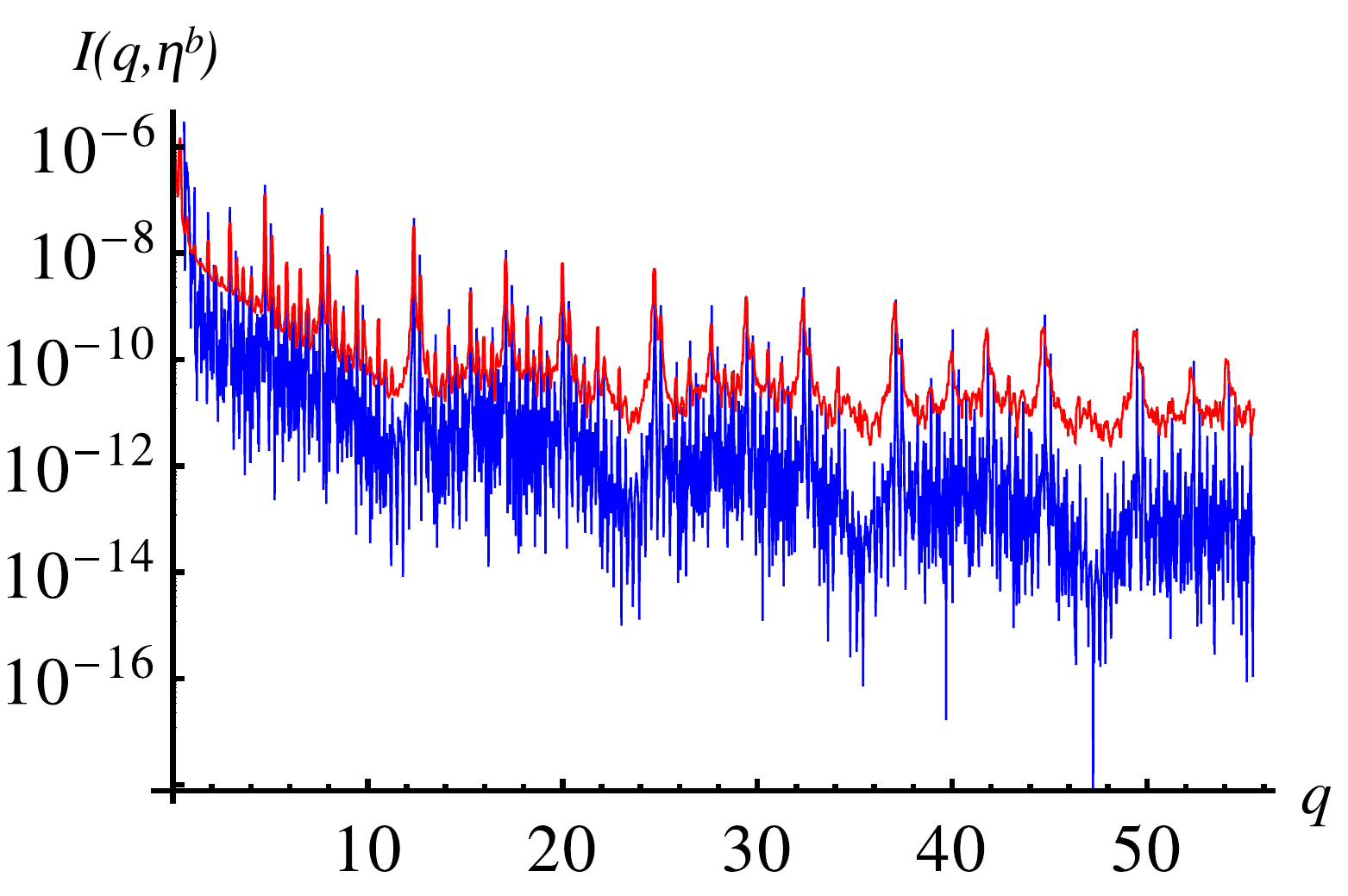}
\protect\protect\caption{(Color online). Experimental diffraction pattern (solid red top curve)
compared with the theoretical diffraction pattern with the inclusion
of the structure factor (solid blue bottom curve) for a Fibonacci
grating with $N=300$ lines and $L=23\;\mu m$ and $S=15\;\mu m$,
$\eta_{3}^{b}=1.875$.}
\label{fig:fib23_15}
\end{figure}

In Fig. \ref{fig:fib23_15} we compare the theoretical diffraction
pattern $I(q,\eta)$ with the experimental
data points for the grating $\eta^{b}_3$. 
The intensity of the experimental points has been rescaled 
to take into account the contribution of the structure factor 
of the grating and the $q$ axis of the experimental
plot has been rescaled in order to compare it with the simulation,
which considers FL's with $S=1$. A similar figure 
is obtained for the case $\eta^{a}_1=17/6$. 
The agreement is very satisfactory: not only the position 
but also the intensity of the diffraction peaks are correctly obtained, 
confirming that our approach is reliable. 

\begin{thebibliography}{99}
\bibitem{limit} This can also be seen by explicit calculation, in
fact it is possible to show that $\eta_{k}=\frac{F_{k+1}\eta_{0}+F_{k}}{F_{k}\eta_{0}+F_{k-1}}$,
where $F_{k}$ are the Fibonacci numbers $(F_{0}=0)$. The above expression
tends to $\tau$ as $k\rightarrow\infty$ due to the fact that $F_{k+1}/F_{k}\rightarrow\tau$
in this limit.
\end{thebibliography}
\end{document}